\documentclass[sigconf]{acmart}

\usepackage{epsfig}
\usepackage{graphicx}
\usepackage{amsmath}
\usepackage{amssymb}
\usepackage{enumerate}
\usepackage{eqnarray}
\usepackage{verbatim}
\usepackage{bbm}
\usepackage{balance}
\usepackage{multirow}
\usepackage{hyperref}
\usepackage{color}
\newcommand{\tabincell}[2]{\begin{tabular}{@{}#1@{}}#2\end{tabular}} 

\def\BibTeX{{\rm B\kern-.05em{\sc i\kern-.025em b}\kern-.08emT\kern-.1667em\lower.7ex\hbox{E}\kern-.125emX}}
    
\copyrightyear{2019}
\acmYear{2019}
\setcopyright{acmcopyright}
\acmConference[MM '19] {Proceedings of the 27th ACM International Conference on Multimedia}{October 21--25, 2019}{Nice, France}
\acmBooktitle{Proceedings of the 27th ACM International Conference on Multimedia (MM'19), Oct. 21--25, 2019, Nice, France}
\acmPrice{15.00}
\acmDOI{10.1145/3343031.3350974}
\acmISBN{978-1-4503-6889-6/19/10}

\begin{document}

\fancyhead{}

\title{A New Benchmark and Approach for \\ Fine-grained Cross-media Retrieval}

\author{Xiangteng He, Yuxin Peng* and Liu Xie}
\affiliation{%
  \institution{Institute of Computer Science and Technology, Peking University, Beijing, China}
  \postcode{100871}
}
\email{pengyuxin@pku.edu.cn}

%
\renewcommand{\shortauthors}{He et al.}

%
\begin{abstract}
Cross-media retrieval is to return the results of various media types corresponding to the query of any media type.
Existing researches generally focus on \textbf{coarse-grained cross-media retrieval}.
When users submit an image of ``Slaty-backed Gull'' as a query, coarse-grained cross-media retrieval treats it as ``Bird'', so that users can only get the results of ``Bird'', which may include other bird species with similar appearance (image and video), descriptions (text) or sounds (audio), such as ``Herring Gull''.
Such coarse-grained cross-media retrieval is not consistent with human lifestyle, where we generally have the fine-grained requirement of returning the exactly relevant results of ``Slaty-backed Gull'' instead of ``Herring Gull''.
However, few researches focus on \textbf{fine-grained cross-media retrieval}, which is a highly challenging and practical task.
Therefore, in this paper, we first construct \textbf{a new benchmark} for fine-grained cross-media retrieval, which consists of \textbf{200} fine-grained subcategories of the ``Bird'', and contains \textbf{4} media types, including image, text, video and audio.
To the best of our knowledge, it is the first benchmark with 4 media types for fine-grained cross-media retrieval.
Then, we propose a uniform deep model, namely \textbf{FGCrossNet}, which simultaneously learns 4 types of media without discriminative treatments.
We jointly consider three constraints for better common representation learning: \emph{classification constraint} ensures the learning of discriminative features for fine-grained subcategories, \emph{center constraint} ensures the compactness characteristic of the features of the same subcategory, and \emph{ranking constraint} ensures the sparsity characteristic of the features of different subcategories.
Extensive experiments verify the usefulness of the new benchmark and the effectiveness of our FGCrossNet.
The new benchmark and the source
code of FGCrossNet will be made available at {\color{magenta}\url{https://github.com/PKU-ICST-MIPL/FGCrossNet_ACMMM2019}}.
\end{abstract}

%
%

\ccsdesc[500]{Information systems~Multimedia and multimodal retrieval}

\ccsdesc[300]{Computing methodologies~Artificial intelligence}

%
\keywords{Fine-grained Cross-media Retrieval, New Benchmark}

%

%
\maketitle

\section{Introduction}
\begin{figure}[!t]
    \begin{center}\includegraphics[width=1\linewidth]{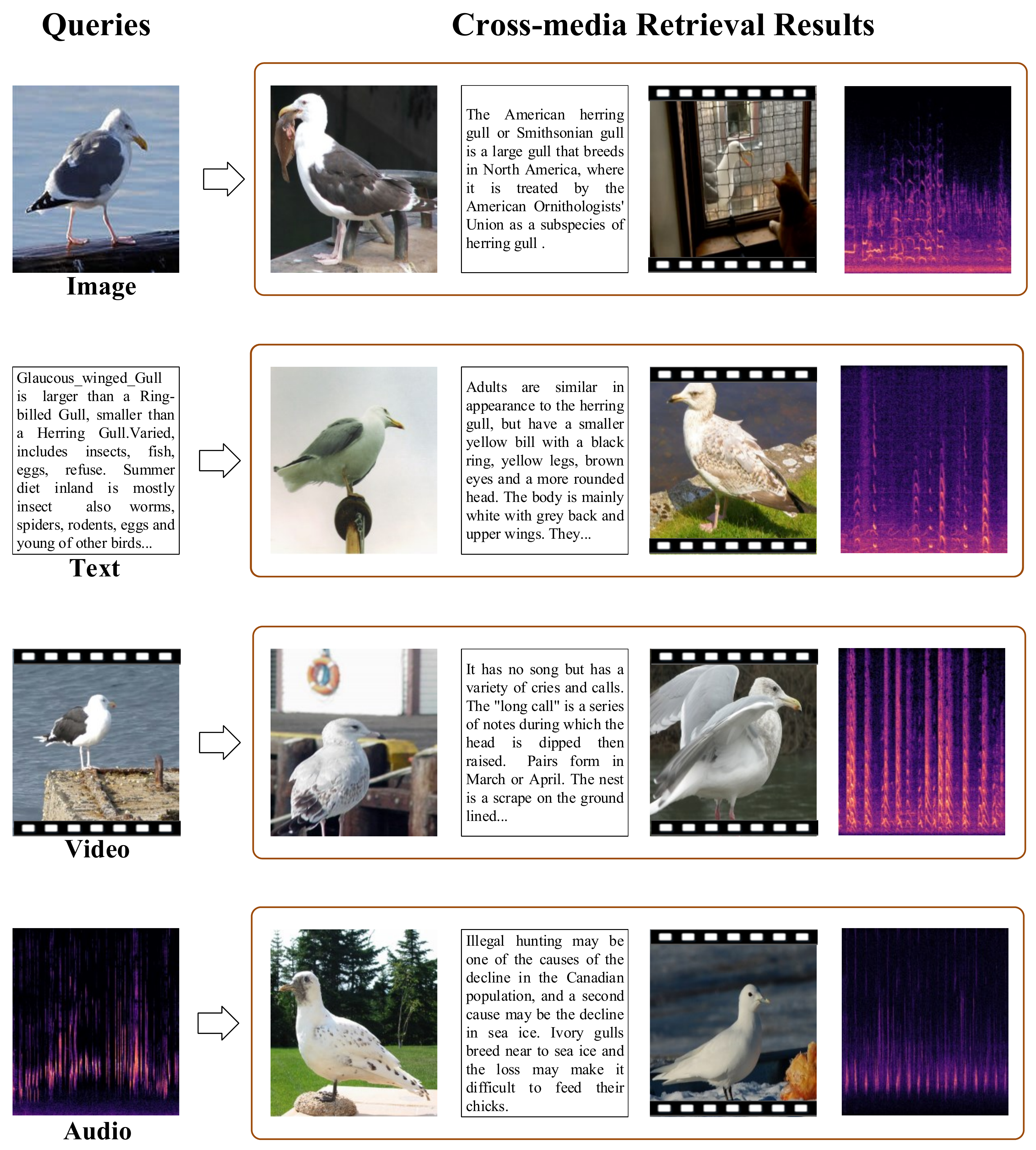}
    \vspace{-6mm}
    \caption{Examples of cross-media retrieval, where the audio data is visualized by spectrogram.}
    \label{cross-media-retrieval}
    \end{center}
\end{figure}

In the era of big data, multimedia data, such as image, text, video and audio, has become the main form of humans knowing the world. Therefore, it is significant to provide an effective multimedia retrieval paradigm to satisfy the requirement of human retrieval.
\textbf{Cross-media retrieval} \cite{peng2018overview} is such an effective retrieval paradigm which users can get the results of various media types by submitting a query of any media type. It has attracted great interests from researchers. Some examples of cross-media retrieval are shown in Figure \ref{cross-media-retrieval}.

\begin{table*}[!t]
  \centering
  \caption{Comparing our new benchmark with other existing coarse-grained cross-media retrieval datasets/benchmarks. }
  \vspace{-3mm}
  \label{stadataset}
  \begin{tabular} {|c|c|c|c|c|c|c|c|}
    \hline
     & Wikipedia & Pascal Sentences & Flickr-30K  & MS-COCO  & NUS-WIDE &PKU XMediaNet & Ours\\
    \hline
    \# images & 2,866 &  1,000 & 31,783 & 123,287 & 269,648 &40,000 & 11,788 \\
    \# texts & 2,866 &  5,000 & 158,915 & 616,435 &5,018 & 40,000 & 8,000 \\
    \# videos & N/A &  N/A & N/A & N/A &N/A & 10,000 & 18,350\\
    \# audios & N/A & N/A & N/A & N/A & N/A & 10,000 & 12,000 \\
    \# categories & 10 & 20 & N/A & 91 & 81 & 200 & 200\\
    fine-grained? & No & No & No & No & No & No & Yes \\
    \hline
  \end{tabular}
\end{table*}

\begin{figure}[!t]
    \begin{center}\includegraphics[width=1\linewidth]{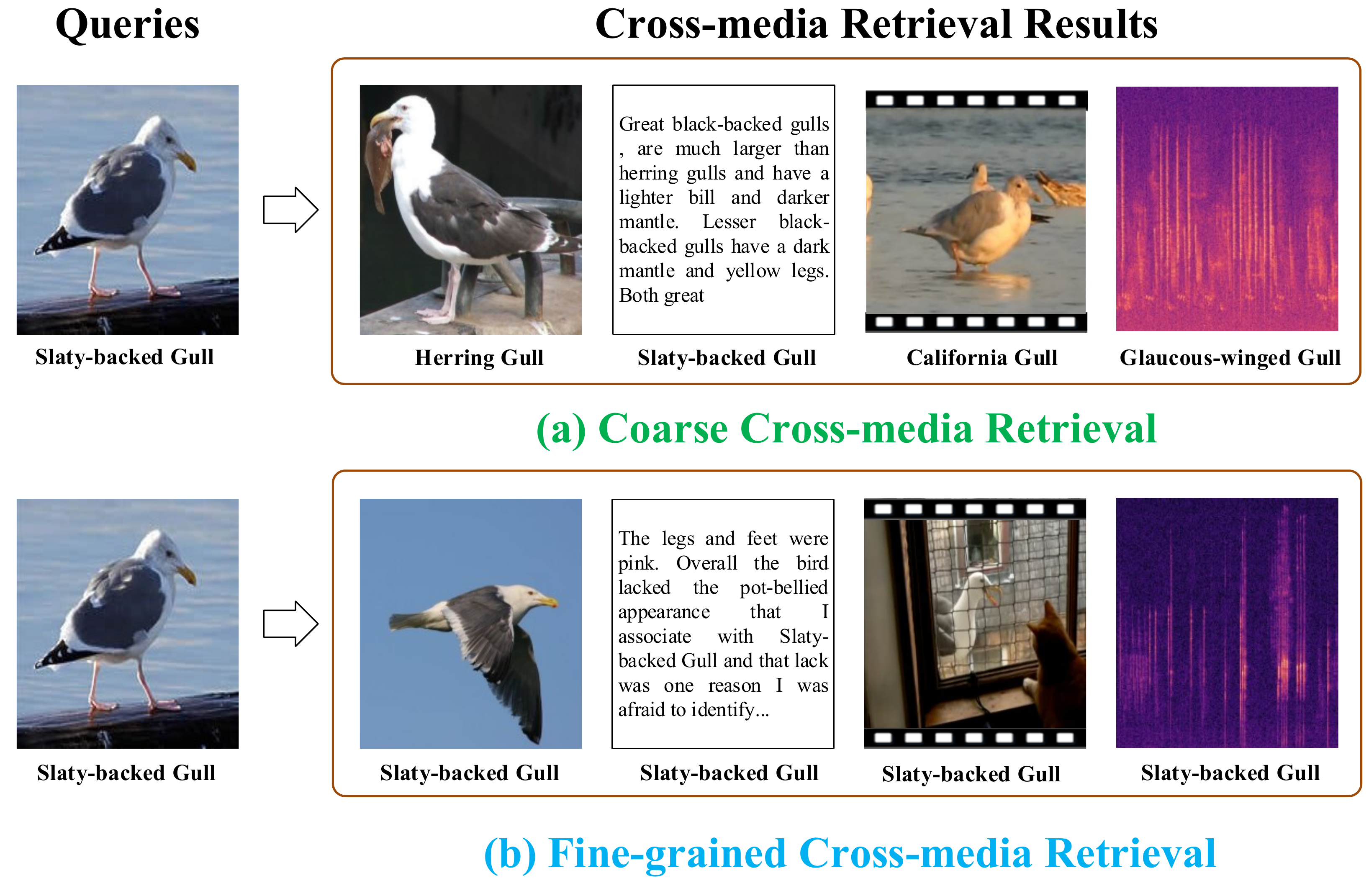}
    \vspace{-6mm}
    \caption{The difference between coarse-grained cross-media retrieval and fine-grained cross-media retrieval, submitting an image of ``Slaty-backed Gull'' as the query.}
    \label{fine-grained}
    \end{center}
\end{figure}

Existing researches mainly concentrate on \textbf{coarse-grained cross-media retrieval}.
As shown in Figure \ref{fine-grained}, when users submit an image of ``Slaty-backed Gull'' as the query, results of various media types will be returned, including image, text, video and audio. In coarse-grained cross-media retrieval, it just treats the image as ``Bird'', so results related to ``Bird'' are returned without further fine-grained consideration.
Thus, the retrieval result may be the image of ``Herring Gull'', which is similar with ``Slaty-backed Gull'' in the global appearance, as shown in Figure \ref{fine-grained} (a). It is hard to distinguish between ``Slaty-backed Gull'' and ``Herring Gull''.
It does not satisfy the \textbf{fine-grained} requirement that we wish to get the results exactly related to ``Slaty-backed Gull'' instead of ``Herring Gull''.
\textbf{Fine-grained cross-media retrieval} is such a paradigm that satisfies the fine-grained retrieval requirement, and returns the exactly related results corresponding to the fine-grained subcategory of the submitted query, as shown in Figure \ref{fine-grained} (b). However, researchers rarely pay attention to this study.

\textbf{Fine-grained cross-media retrieval} has three challenges: (1) \textbf{Few datasets} - Existing cross-media datasets are mainly constructed for coarse-grained cross-media retrieval with coarse-grained categories or semantics, but few datasets are available to explore the fine-grained cross-media retrieval.
(2) \textbf{Heterogeneity gap} - Variant types of media have inconsistent distributions and feature representations, which makes the cross-media retrieval quite challenging. 
(3) \textbf{Small inter-class variance} - Similar subcategories that belong to the same basic-level genus may have the similar global appearance (image or video), similar textual descriptions (text) and similar sounds (audio), which causes it difficult to discriminate similar fine-grained subcategories.

To address these issues, we first construct a new benchmark for fine-grained cross-media retrieval, and then propose a uniform deep model (FGCrossNet) to learn the common representations for 4 types of media simultaneously. Comprehensive experimental results and analyses on the new benchmark verify its usefulness and the effectiveness of our FGCrossNet.
We summarize the contributions of this paper as follows:
\begin{enumerate}[(I)]
\item
\textbf{A new benchmark for fine-grained cross-media retrieval} is constructed. It has 3 advantages: (1) \textbf{Species diversity} - It consists of \emph{200} fine-grained subcategories of the ``Bird'', and contains \emph{4} media types, including \emph{image, text, video and audio}. To the best of our knowledge, it is the largest benchmark with the most media types for fine-grained cross-media retrieval.
(2) \textbf{Domain diversity} - Its data is collected from variant sources (domains), which causes that even the data of the same media type has different attributes and distributions, as well as increases the challenge of fine-grained cross-media retrieval.  
(3) \textbf{Availability} - It will be released publicly for the researchers to promptly
evaluate their methods on the new benchmark, which encourages further studies on fine-grained cross-media retrieval. 
\item
\textbf{A uniform deep model for fine-grained cross-media retrieval} is proposed, namely FGCrossNet, which simultaneously learns 4 types of media without discriminative treatments.
We jointly consider three constraints for better common representation learning: \emph{classification constraint} ensures the learning of discriminative features for fine-grained subcategories, \emph{center constraint} ensures the compactness characteristic of the features of the same subcategory, and \emph{ranking constraint} ensures the sparsity characteristic of the features of different subcategories.
\end{enumerate}


\section{A New Benchmark}
\label{newbenchmark}
There are already several datasets/benchmarks for coarse-grained cross-media retrieval, and their statistical information is shown in Table \ref{stadataset}.
Concretely, Rasiwasia et al. construct the most widely-used cross-media dataset, namely Wikipedia \cite{rasiwasia2010new}, which contains 2,866 image/text pairs from 10 coarse-grained categories, such as ``History'' and ``Warfare''.
Rashtchian et al. select 1,000 images from Pascal VOC 2008 dataset \cite{pascal-voc-2008}, and annotate each of them with 5 sentences to construct the Pascal Sentences dataset \cite{rashtchian2010collecting}.
Later on, some large scale cross-media datasets are constructed to boost the development of coarse-grained cross-media retrieval, such as Flickr-30K \cite{young2014image} and MS-COCO \cite{lin2014microsoft}.
The textual information in these datasets is represented by sentences or articles.
Chua et al. construct the NUS-WIDE dataset \cite{chua2009nus}, which collects 269,648 images of 81 coarse-grained categories from the websites. There are 5,018 unique tags, which represent the textual information of the corresponding images.
These datasets have the same limitation that they only contain \textbf{2 media types}, i.e. image and text.

For comprehensive evaluation and boosting the development of coarse-grained cross-media retrieval, Peng et al. construct the PKU XMediaNet \cite{peng2018overview}, which is the largest cross-media dataset with up to 5 media types, including image, text, video, audio and 3D model.
It contains 100,000 samples from 200 coarse-grained categories. Its categories are selected from WordNet \footnote{wordnet.princeton.edu/}, and cover 47 species of animals such as ``Bird'' and ``Dog'', as well as 153 types of artifacts, such as ``Airplane'' and ``Car''. However, the above datasets only contain basic-level \textbf{coarse-grained} categories, so that they cannot satisfy the \textbf{fine-grained} retrieval requirement.

Therefore, in this paper, we construct a new benchmark for fine-grained cross-media retrieval, which consists of \emph{\textbf{4 media types}}, including image, text, video and audio, as well as contains \emph{\textbf{200 fine-grained subcategories}} that belong to the coarse-grained category of ``Bird''. From Table \ref{stadataset}, we can see that our new benchmark is the largest benchmark with the most media types for fine-grained cross-media retrieval. In the following paragraphs, we will introduce it in details from three aspects: collection and properties of the new benchmark, as well as the fine-grained cross-media retrieval task.

\subsection{Collection}
\label{collection}
We collect data with variant media types, including image, text, video and audio, to construct the new benchmark for fine-grained cross-media retrieval. Inspired by the works in fine-grained visual categorization \cite{wah2011caltech,zhu2018fine}, we construct the new benchmark consisting of 200 fine-grained subcategories of ``Bird''. Researchers have built the image and video datasets consisting of 200 bird species with the same taxonomy, namely CUB-200-2011 \cite{wah2011caltech} and YouTube Birds \cite{zhu2018fine}. So we build the new benchmark on the basis of these two datasets, and directly use them as the image and video data. We first briefly introduce the two datasets as follows:

\textbf{CUB-200-2011} \cite{wah2011caltech} is the most widely-used fine-grained image classification \cite{he2019and,he2017fine} dataset, including 11,788 images of 200 subcategories belonging to the same basic-level coarse-grained category of ``Bird''. It is divided as follows: training set contains 5,994 images, and testing set contains 5,794 images.
Each image has detailed annotations: an image-level subcategory label, a bounding box of the object, 15 part locations and 312 binary attributes.

\textbf{YouTube Birds} \cite{zhu2018fine} is a new fine-grained video dataset, including 18,350 videos of 200 subcategories belonging to the same basic-level coarse-grained category of ``Bird''. Its taxonomy is the same with CUB-200-2011 dataset, and its video instances are collected from YouTube. The duration of each video is no more than 5 minutes. It is divided as follows: training set contains 12,666 videos, and testing set contains 5,684 videos.

Besides, we need to collect text and audio data.
Since they are easily available on the Internet, we select some professional websites as our data sources, as shown in Table \ref{datasource}. In the following paragraphs, we introduce the collection process in the two aspects: collecting and cleaning.

\subsubsection{\textbf{Collecting}}
\label{collecting}
\indent

\textbf{Text Collecting}:
Wikipedia \footnote{www.wikipedia.org/} is the largest free online encyclopedia, created and edited by volunteers around the world. From Wikipedia, we can easily get the corresponding textual descriptions by submitting the names of fine-grained subcategories as the query keywords. Note that the names of fine-grained subcategories are same with CUB-200-2011 dataset.
From Wikipedia, we obtain the text data of 200 subcategories. However, the text instances of each subcategory are not enough.
To get more text data, we apply two strategies:
(1) \emph{More encyclopedia websites} - Except the Wikipedia, we obtain the text data from the other 11 professional websites, such as All About Birds, Audubon, and Animal Spot, as shown in Table \ref{datasource}.
(2) \emph{More query keywords} - A lot of the bird species have scientific names or aliases, which can be taken as the query keywords for getting more text data. For example, ``Black-footed Albatross'' has the scientific name as ``Phoebastria Nigripes Audubon''.

\textbf{Audio Collecting}: 
Same with text collecting, we also select the professional audio websites as the audio data sources, such as xeno-canto \footnote{www.xeno-canto.org} and Bird-sounds \footnote{www.bird-sounds.net/} which share sounds of different bird species from around the world. To get more audio data, we also apply the two strategies as text collecting: \emph{more professional audio websites}, totally 7 websites as shown in Table \ref{datasource}, and \emph{more query keywords}.

\begin{table}[!t]
  \centering
  \caption{Data sources for text and audio. }
  \vspace{-3mm}
  \label{datasource}
  \begin{tabular} {|c|c|}
    \hline
    Data & Data Sources \\
    \hline
    Text & \tabincell{c}{(1) www.wikipedia.org (2)  www.allaboutbirds.org \\ (3) www.audubon.org (4) birdsna.org \\(5) birds.fandom.com (6) nhpbs.org (7) ebird.org \\(8) mnbirdatlas.org (9) sites.psu.edu \\ (10) www.birdwatchersdigest.com (11) folksread.com \\ (12) neotropical.birds.cornell.edu}\\
    \hline
    Audio & \tabincell{c}{(1) www.xeno-canto.org (2) www.bird-sounds.net \\ (3) www.findsounds.com (4) freesound.org \\ (5) www.macaulaylibrary.org (6) avibase.bsc-eoc.org \\ (7) soundcloud.com}\\
    \hline
  \end{tabular}
\end{table}

 
\subsubsection{\textbf{Cleaning}}
\label{cleaning}
\indent

\textbf{Text Cleaning}:
There are some noises in the collected data. We first remove the web page links from the text data, and then divide one textual article by paragraphs. After that each paragraph is taken as a text instance, which is the final text data.
Since these text data is collected from the professional encyclopedia data, they have been well labeled.

\begin{figure*}[!t]
    \begin{center}\includegraphics[width=0.95\linewidth]{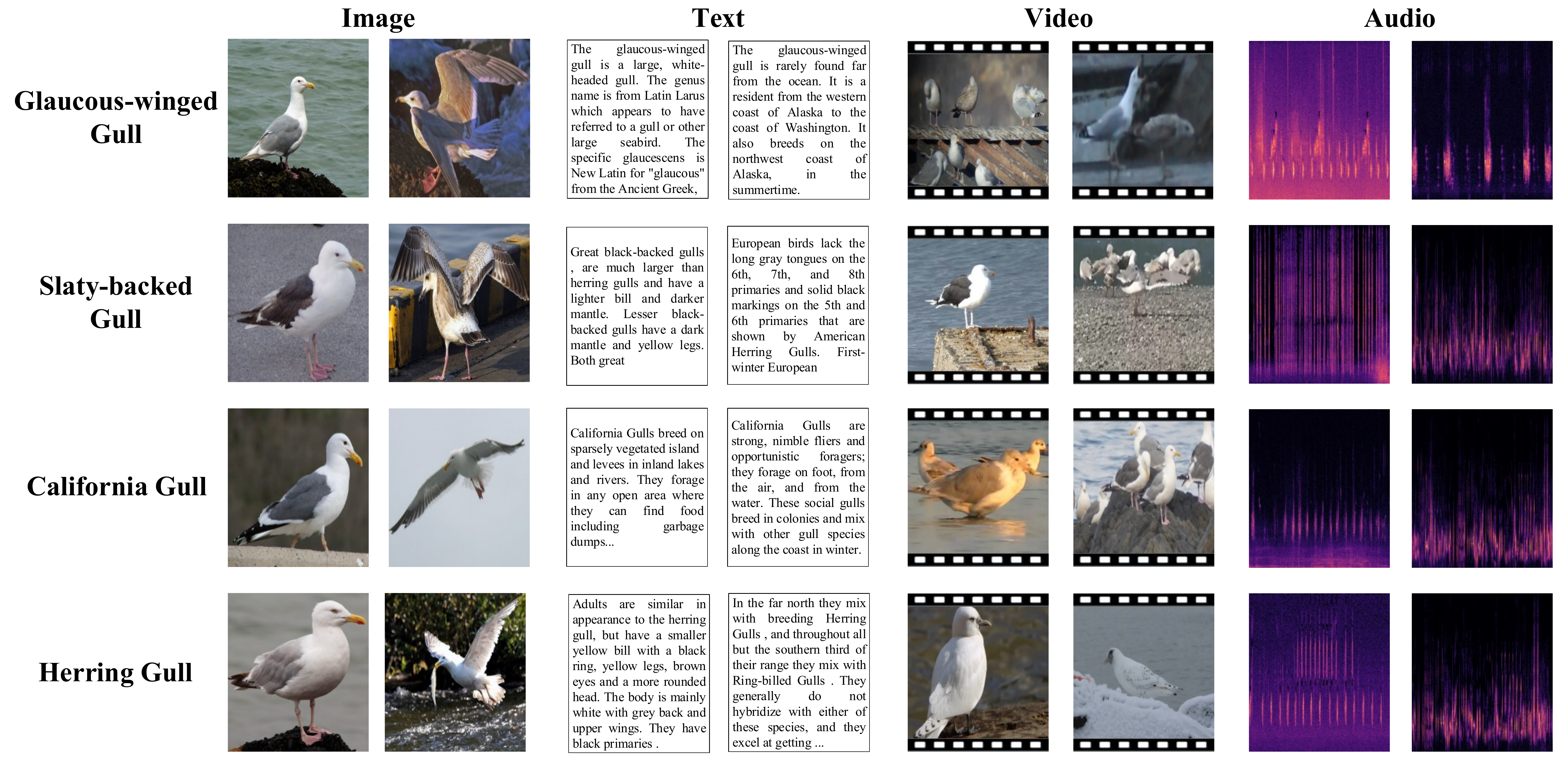}
    \vspace{-3mm}
    \caption{Some examples of 4 similar subcategories in our new benchmark, where the audio data is visualized by spectrogram.}
    \label{dataset}
    \end{center}
    \vspace{-4mm}
\end{figure*}

\textbf{Audio Cleaning}:
Since the durations of some collected audio instances are too long, e.g. more than an hour, we divide the audio into several parts to get more audio instances.
However, the division causes some audio instances have no sounds of the bird species, so we ask human annotators to delete these audio instances. Note that some audio instances contain other sounds, such as the sounds of humans or wind, which increases the challenge of fine-grained cross-media retrieval.

Through collecting and cleaning, we get the final data for fine-grained cross-media retrieval, some examples of the fine-grained cross-media data are shown in Figure \ref{dataset}.

\subsection{Properties}
\label{properties}

\subsubsection{\textbf{Scale}}
\label{scale}
\indent

From Table \ref{stadataset}, we can see that our new benchmark contains 4 media types, is only inferior to PKU XMediaNet dataset \cite{peng2018overview}, which contains the media type of 3D model additionally.
The other cross-media datasets only consist of 2 media types, i.e. image and text. Besides, the scale of each media type in the new benchmark is large, i.e. 11,788 images, 8,000 texts, 18,350 videos and 12,000 audios. For text, there are 40 instances of each subcategory. For audio, there are 60 instances of each subcategory.

\subsubsection{\textbf{Diversity}}
\label{diversity}
\indent

\textbf{Species Diversity}
The newly constructed benchmark contains 200 subcategories that are corresponding to 200 bird species. This property makes the new benchmark be the largest with the most media types for fine-grained cross-media retrieval. Similar fine-grained subcategories bring the challenge of small inter-class variance: They have similar global appearance (image or video), similar textual descriptions (text) and similar sounds (audio), which makes it hard to discriminate similar subcategories. For example, in Figure \ref{dataset}, even the image examples belong to different subcategories, they look similarly in global appearance.

\textbf{Domain Diversity}
All the data is collected from different sources (domains) with variant qualities, which causes the shifts among the data distributions, and increases the challenge of fine-grained cross-media retrieval. For the images and videos, they are variant in resolution, color, view, illumination. For texts, they are variant in length. For audios, they are variant in length and background sound.
The durations of audios are variant from 1 second to 2,000 seconds. Some audios contain not only the bird sounds, but also some other sounds, such as humans and wind.

\subsection{Fine-grained Cross-media Retrieval}
To demonstrate the usefulness of our newly constructed benchmark, we conduct the following two tasks for evaluating the fine-grained cross-media retrieval performance of different methods, namely \emph{bi-modality fine-grained cross-media retrieval} and \emph{multi-modality fine-grained cross-media retrieval}, following \cite{peng2018overview}.

\textbf{Bi-modality fine-grained cross-media retrieval}: The query is one instance of any media type, and the retrieval results are instances of the other one media type. For example, if the query is an image of ``Slaty-backed Gull'', the results can be text instances of ``Slaty-backed Gull'', which is denoted as ``I $\to$ T''. There are totally 12 retrieval tasks, including ``I $\to$ T'', ``I $\to$ V'', ``I $\to$ A'', ``T $\to$ I'', ``T $\to$ V'', ``T $\to$ A'', ``V $\to$ I'', ``V $\to$ T'', ``V $\to$ A'', ``A $\to$ I'', ``A $\to$ T'' and ``A $\to$ V''.

\textbf{Multi-modality fine-grained cross-media retrieval}: The qu-ery is one instance of any media type, and the retrieval results are instances of all media types. For example, if the query is an image of ``Slaty-backed Gull'', the results will be instances of ``Slaty-backed Gull'' in the types of image, text, video and audio, which is denoted as ``I $\to$ all''.
There are totally 4 retrieval tasks, including ``I $\to$ all'', ``T $\to$ all'', ``V $\to$ all'' and ``A $\to$ all''.

\begin{figure*}[!t]
    \begin{center}\includegraphics[width=0.85\linewidth]{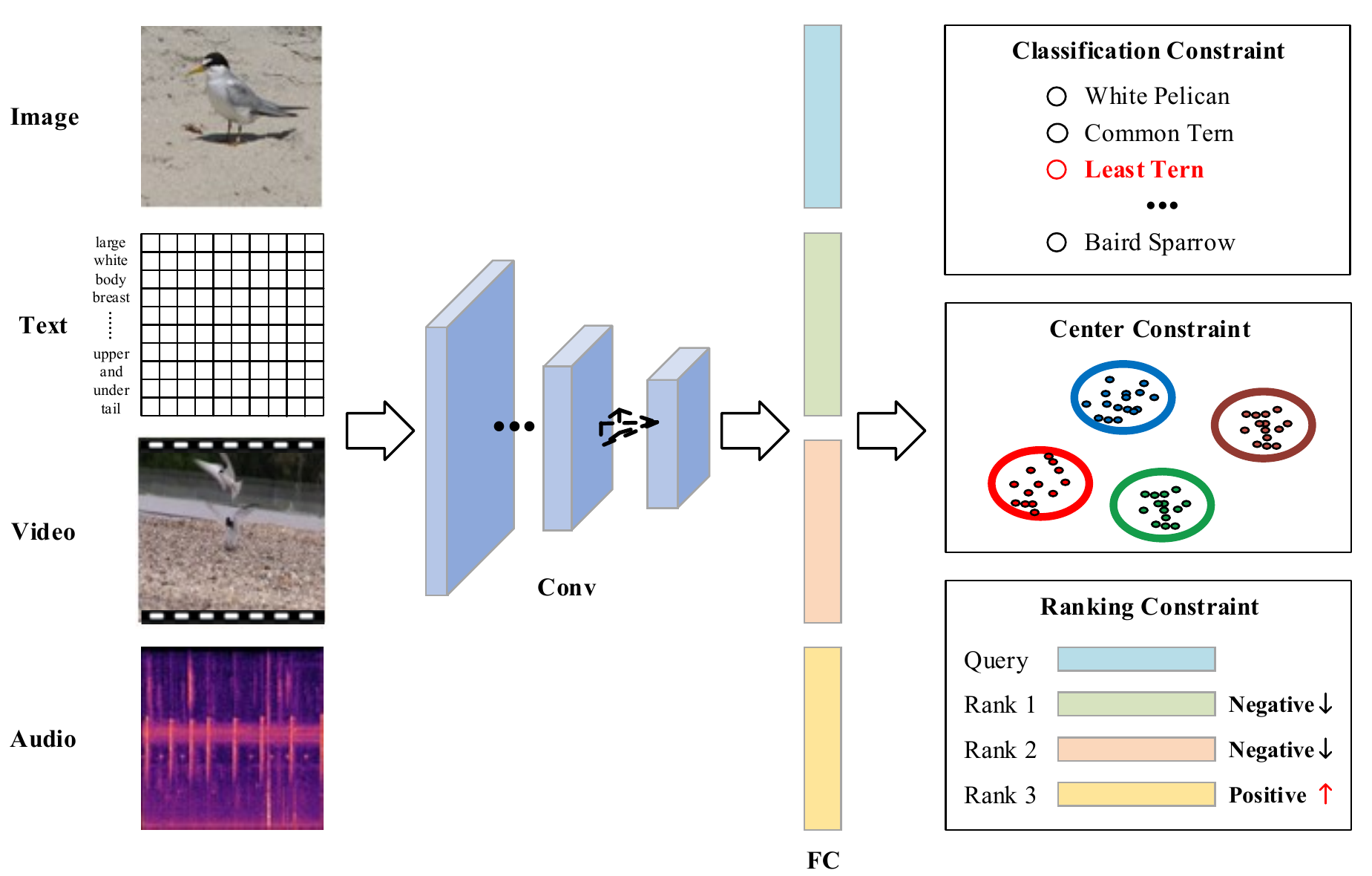}
    \vspace{-3mm}
    \caption{Overview of our FGCrossNet.}
    \label{framework}
    \end{center}
\end{figure*}

\section{Our Approach}
\label{ourapproach}
To demonstrate the usefulness of our newly constructed benchmark, we also propose a uniform deep model for fine-grained cross-media retrieval, namely FGCrossNet. 
In the following paragraphs, we introduce it in the aspects of \emph{network architecture}, \emph{data preprocessing}, \emph{loss function}, \emph{training} and \emph{retrieval}.

\subsection{Network Architecture}
\label{netarchitecture}
Existing cross-media retrieval methods generally deal with different media data through different network streams, which causes some issues:
(1) \emph{Architecture complexity} - Different media data maybe processed by different types of networks. For example, image is generally processed by convolutional neural networks, such as ResNet \cite{he2016deep}, while text may be processed by LSTM \cite{hochreiter1997long}.
Thus, the final network architecture combines different types of networks, which is a highly complex deep model.
(2) \emph{Training difficulty} - Since the network architecture is complex, its training certainly will be difficult, which causes it hard to reproduce the method.
To simplify the architecture complexity and reduce the training difficulty, we propose a uniform deep model, which adopts the same architecture to simultaneously learn 4 media data without discriminative treatments.
Its architecture is shown in Figure \ref{framework}.
We adopt ResNet50 \cite{he2016deep} as our basic deep model. To achieve better performance, we make some modifications: take $448 \times 448$ as input size, and follow an average pooling layer with kernel size $14$ and stride $1$ after the last convolutional layer.
It is noted that it can be replaced by any other state-of-the-art deep convolutional neural networks, such as AlexNet \cite{krizhevsky2012imagenet} and VGGNet \cite{simonyan2014very}.

\subsection{Data Preprocessing}
\label{datapreprocessing}
To take different media data as the input of our FGCrossNet, we need to conduct data preprocessing firstly. For image, there is no need to do any preprocessing. For video, we draw 25 uniformly-spaced frames of each video as the video data. For audio, we apply Short-Time Fourier Transformation \cite{grochenig2013foundations} to generate spectrogram for each audio instance following \cite{wu2016multi}, so that our FGCrossNet can deal with the audio data.
We generate the spectrogram for each audio by librosa\footnote{librosa.github.io/librosa/core.html}, and set the size of the output spectrogram image as 448 $\times$ 448 in our experiments, which is independent of the audio's length. The examples of spectrograms are shown in Figure \ref{dataset}. 

\begin{figure*}[!t]
    \begin{center}\includegraphics[width=0.8\linewidth]{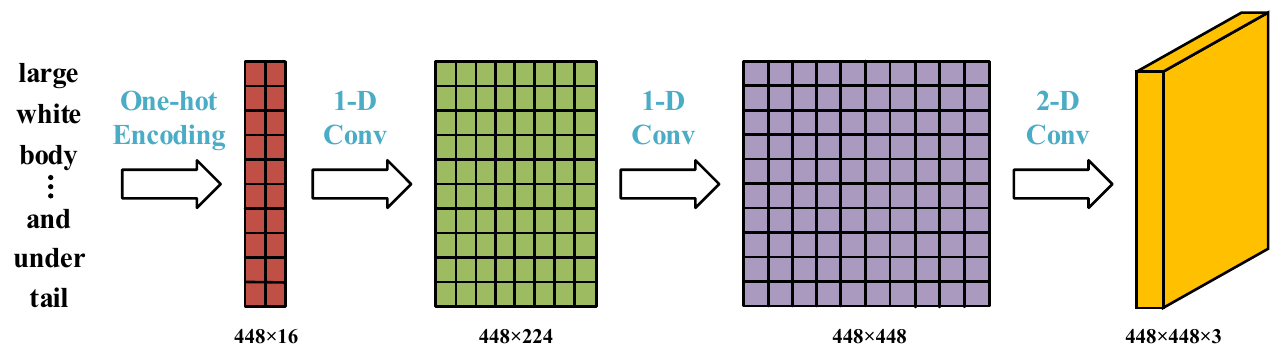}
    \vspace{-3mm}
    \caption{Overview of text processing.}
    \label{textcnn}
    \end{center}
\end{figure*}
For text, to satisfy the input format of our FGCrossNet, we design a text processing approach, the whole process is shown in Figure \ref{textcnn}. Given a text, first we follow \cite{conneau2016very} to convert it into a vector of size $n \times d$ by quantizing each character using one-hot encoding, and the character embedding size is 16. Besides, we fix the max character number of the text is 448, so the vector size is $448 \times 16$. If the character number of the text is less than 448, we pad with zeros to the rows of the vector. If the character number is larger, then the text is truncated. All text data in our new benchmark contains less than 448 characters, so that no information will be lost. 
Then, we apply two 1D convolutional layers of $224$ and $448$ convolutions with size $3$, padding $1$ and stride $1$ respectively, so the output is $448 \times 448$. Finally, we apply a 2D convolutional layer of $3$ convolutions with size $3$, padding $1$ and stride $1$, so the final output is $448 \times 448 \times 3$, which is taken as the input of FGCrossNet. 
Furthermore, we apply position shift \cite{zheng2017dual} to augment the text data for better learning of our FGCrossNet.

\vspace{-3mm}
\subsection{Loss Function}
\label{lossfunction}
We design a new loss function to drive the learning of our FGCrossNet, which jointly considers three constraints for better common representation learning: \emph{classification constraint} ensures the learning of discriminative features for fine-grained subcategories, \emph{center constraint} ensures the compactness characteristic of the features of the same subcategory, and \emph{ranking constraint} ensures the sparsity characteristic of the features of different subcategories.
The new designed loss function is defined as follows:
\begin{gather}
\label{lossfunc}
\mathcal{L} = \mathcal{L}_{cls}+ \mathcal{L}_{cen} + \mathcal{L}_{rank}
\end{gather}
where the three items denote classification constraint, center constraint and ranking constraint respectively.

\subsubsection{\textbf{Classification Constraint}}
\label{classification}
\indent

We apply cross-entropy loss function as classification constraint to drive our FGCrossNet to have the ability of distinguishing one subcategory from other similar subcategories. For example, distinguish ``Slaty-backed Gull'' from ``Herring Gull'', as shown in Figure \ref{dataset}.
The classification constraint $\mathcal{L}_{cls}$ is defined as follows:
\begin{gather}
\label{cls}
\mathcal{L}_{cls} = \frac{1}{N_I} \sum_{k=1}^{N_I}l(x_k^I,y_k^I) + \frac{1}{N_T} \sum_{k=1}^{N_T}l(x_k^T,y_k^T) + \nonumber\\ 
\frac{1}{N_V} \sum_{k=1}^{N_V}l(x_k^V,y_k^V) + \frac{1}{N_A} \sum_{k=1}^{N_A}l(x_k^A,y_k^A)
\end{gather} 
where $l(x_k, y_k)$ is the cross-entropy loss function, $I$, $T$, $V$ and $A$ denote the media types of image, text, video and audio respectively. Take image as example, $N_I$ denotes the number of image data in the training set, $y_k^I$ denotes the label of the $k$-th image data, $x_k^I$ denotes the feature of the $k$-th image data, which is the output of FC layer of FGCrossNet, as shown in Figure \ref{framework}. It is noted that since we draw 25 uniformly-spaced frames of each video, $N_V$ denotes the number of all video frames in the training set.

\subsubsection{\textbf{Center Constraint}}
\label{center}
\indent

In Equation (\ref{lossfunc}), the second item $\mathcal{L}_{cen}$ denotes the center constraint, and its definition is as follows:
\begin{gather}
\label{cen}
\mathcal{L}_{cen} = \frac{1}{2}\sum_{k=1}^{N}||x_k - c_{y_k}||^2_2
\end{gather}

To achieve better performance of fine-grained cross-media retrieval, the features of the same subcategory should be adjacent in the common space, which is to minimize the intra-class variance and reduce the domain shift.
Inspired by clustering, we drive the learning of our FGCrossNet through minimizing the distance of the feature to its subcategory center.
In Equation (\ref{cen}), $x_k$ denotes the feature of the $k$-th training data, which may be any media type. In center constraint, we do not discriminate which media type $x_k$ belongs to, but treat all media data equally, because we focus the compactness characteristic of the features of the same subcategory. Therefore, $N$ denotes the number of all training media data, and $c_{y_k}$ denotes the $y_k$ subcategory center, which is updated every batch in the training phase, and calculated by the features of all the media data of $y_k$ in a batch.

\subsubsection{\textbf{Ranking Constraint}}
\label{rankingconstraint}
\indent 

In Equation (\ref{lossfunc}), the third item $\mathcal{L}_{rank}$ denotes the ranking constraint, and its definition is as follows:

\begin{gather}
\mathcal{L}_{rank} = \sum_{i, j, k}^{N}(d(x_i, x_j)^2 - d(x_i, x_k)^2 + \alpha_1)_+\nonumber\\ 
+ \sum_{i, j, k, l}^{N}(d(x_i, x_j)^2 - d(x_l, x_k)^2 + \alpha_2)_+
\label{rankloss}
\end{gather}
\begin{table*}[!t]
  \centering
  \caption{The MAP scores of bi-modality fine-grained cross-media retrieval of our FGCrossNet compared with existing methods on all 4 media types, including image, text, video and audio. }
  \vspace{-3mm}
  \label{expall}
  \begin{tabular} {|c|c|c|c|c|c|c|c|c|c|c|c|c|c|}
    \hline
    Methods & I$\to$T & I$\to$A& I$\to$V& T$\to$I& T$\to$A& T$\to$V& A$\to$I& A$\to$T& A$\to$V& V$\to$I& V$\to$T& V$\to$A & Average\\
    \hline
    \textbf{Our FGCrossNet}  & \textbf{0.210}  &\textbf{0.526}  &\textbf{0.606} &\textbf{0.255}&\textbf{0.181}  &\textbf{0.208} &\textbf{0.553} &\textbf{0.159}   &\textbf{0.443}&  \textbf{0.629}&\textbf{0.195} &\textbf{0.437} &\textbf{0.366} \\
    MHTN \cite{huang2018mhtn} &0.116& 0.195 &0.281  &0.124  &0.138  &0.185  &0.196  &0.127  &0.290  &0.306  &0.186  &0.306 &0.204  \\
    ACMR \cite{wang2017adversarial} &0.162  &0.119  &0.477  &0.075  &0.015  &0.081  &0.128  &0.028  &0.068  &0.536  &0.138  &0.111  &0.162\\
    JRL \cite{zhai2014learning} &0.160  &0.085& 0.435&  0.190&  0.028&  0.095&  0.115&  0.035&  0.065&  0.517&  0.126&  0.068&  0.160 \\
    GSPH \cite{mandal2017generalized} & 0.140  &0.098 &0.413 &0.179 &0.024 &0.109 &0.129 &0.024 &0.073 &0.512 &0.126 &0.086 &0.159\\
    CMDN \cite{peng2016cross} & 0.099  &0.009  &0.377 & 0.123  &0.007  &0.078 & 0.017 & 0.008  &0.010  & 0.446  &0.081  &0.009  &0.105 \\
    SCAN \cite{lee2018stacked} & 0.050& -&- &0.050 & -&- &- &- &- &- &- &-  & 0.050 \\
    GXN \cite{gu2018look} & 0.023 &- &-& 0.035 &- &- &- & -& -&- &- &-  &  0.029 \\
    \hline
  \end{tabular}
\end{table*}

Since center constraint is to minimize the intra-class variance, ranking constraint is to maximize the inter-class variance.
We apply the quadruplet loss function\cite{chen2017beyond} to drive our FGCrossNet to lead the feature outputs of different subcategories to be more dissimilar than those of same subcategory.
In Equation (\ref{rankloss}), $x$ denotes the training media data. It is noted that $x_i$, $x_j$, $x_k$ and $x_l$ denote 4 input instances of 4 media types respectively.
There are two constraints among these 4 instances: (1) They must be different media types, i.e. one image, one text, one video and one audio. (2) They must belong to 3 subcategories, where two of the 4 instances are from the same subcategory, and the other two are from the left two subcategories respectively.
For example, $x_i$, $x_j$, $x_k$ and $x_l$ denote image data, text data, video data, and audio data respectively, $x_i$ and $x_j$ are from the subcategory of ``Slaty-backed Gull'', $x_k$ is from ``California Gull'' and $x_l$ is from ``Herring Gull'', which constitute the quadruplet. More details of the quadruplet setting will be introduced in Section \ref{training}.
Their variance is measured by L2 distance, which is denoted as $d()$. $\alpha_1$ and $\alpha_2$ denote the margin thresholds, which are to determine the balance of two terms of Equation (\ref{rankloss}). We set them to $1$ and $0.5$ as same as \cite{chen2017beyond}.

\subsection{Training}
\label{training}

\subsubsection{\textbf{Input}}
\label{input}
\indent

Instead of taking only one instance as input, we take 4 instances as input at the same time, and the 4 instances are from image, text, video and audio respectively. Besides, to calculate the ranking constraint, we restrict that the 4 instances belong to 3 subcategories, which means two of them belong to the same subcategory. It is noted that this setting has nothing to do with the media types, so that the two instances belong to the same subcategory can be of any media type, and they are randomly selected to be $x_i$ and $x_j$.

\subsubsection{\textbf{Training Strategy}}
\label{trainingstrategy}
\indent

Since the inputs of FGCrossNet are images (image, video and audio) or image-like matrix (text), we first only take the image data as input to fine-tune our FGCrossNet, which is pre-trained on ImageNet dataset \cite{deng2009imagenet}. Then, we take 4 instances of 4 media types as inputs, as described in Section \ref{input}, to fine-tune our FGCrossNet by minimizing the new proposed loss function $\mathcal{L}$ with classification and center constraints first, then with all the three constraints. In the fine-tune phase, the learning rate starts with 0.001, and decreases by 0.5 for every 3 epochs.

\subsection{Retrieval}
When retrieval, we extract the outputs of FC layer in our FGCrossNet as the common representations for 4 media types. Then, we apply cosine distance to measure the similarities across different media data. Finally, we return the results based on the similarities.

\section{Experiment}
\label{experiment}
To demonstrate the usefulness of the new benchmark and the effectiveness of our FGCrossNet, we conduct fine-grained cross-media retrieval task on the newly constructed benchmark and compare with state-of-the-art methods.

\subsection{Data and Evaluation Metric}
\subsubsection{\textbf{Data Division}}
\indent

For image and video, we follow the division settings of the original datasets. For image, the training set contains 5,994 images, and the testing set contains 5,794 images. For video, the training set contains 12,666 videos, and the testing set contains 5,684 videos. For text, the training and testing sets contain 4,000 texts respectively. For audio, both of the training and testing sets contain 6,000 audios.

\subsubsection{\textbf{Evaluation Metric}}
\indent

Following \cite{peng2018overview}, we apply the mean average precision (MAP) score to evaluate the fine-grained cross-media retrieval performance. We first calculate average precision (AP) score for each query, and then calculate their mean value as MAP score.

\begin{table}[!t]
  \centering
  \caption{The MAP scores of multi-modality fine-grained cross-media retrieval of our FGCrossNet compared with existing methods on all 4 media types, including image, text, video and audio. }
  \vspace{-3mm}
  \label{exptoall}
  \scalebox{0.93}{
  \begin{tabular} {|c|c|c|c|c|c|}
    \hline
    Methods & I$\to$All & T$\to$All & V$\to$All & A$\to$All & Average\\
    \hline
    \textbf{Our FGCrossNet}  &\textbf{0.549}& \textbf{0.196}& \textbf{0.416}& \textbf{0.485}& \textbf{0.412} \\
    MHTN \cite{huang2018mhtn} & 0.208
&0.142  &0.237  &0.341 & 0.232   \\
    GSPH \cite{mandal2017generalized} &0.387  & 0.103&0.075 &0.312 &0.219  \\
    JRL \cite{zhai2014learning} & 0.344 &0.080  &0.069  &0.275 &0.192  \\
    CMDN \cite{peng2016cross} &0.321  &0.071  &0.016  &0.229 &0.159  \\
    ACMR \cite{wang2017adversarial} &0.245  &0.039  &0.041  &0.279 &0.151  \\   
    \hline
  \end{tabular}}
\end{table}
\begin{table*}[!t]
  \centering
  \caption{Effect of each constraint in our FGCrossNet. }
  \vspace{-3mm}
  \label{ablation}
  \scalebox{0.98}{
  \begin{tabular} {|c|c|c|c|c|c|c|c|c|c|c|c|c|c|}
    \hline
    Methods & I$\to$T & I$\to$A& I$\to$V& T$\to$I& T$\to$A& T$\to$V& A$\to$I& A$\to$T& A$\to$V& V$\to$I& V$\to$T& V$\to$A & Average\\
    \hline
    Classification Constraint  &   0.132  & 0.485& 0.579  & 0.181 & 0.126 &   0.146 & 0.514 & 0.100 & 0.410 & 0.597  & 0.126  &  0.396  & 0.316 \\
    +Center Constraint   &  0.195 &\textbf{0.540} &0.596  &0.240& 0.176&  0.193&  \textbf{0.562}& 0.150 &0.439& 0.616&  0.174&  0.432&  0.359 \\
    +Ranking Constraint   & \textbf{0.210}  &0.526  &\textbf{0.606} &\textbf{0.255}&\textbf{0.181}  &\textbf{0.208} &0.553  &\textbf{0.159}   &\textbf{0.443}&  \textbf{0.629}&\textbf{0.195} &\textbf{0.437} &\textbf{0.366} \\
    \hline
  \end{tabular}}
\end{table*}

\subsection{Compared Methods}
We compare our FGCrossNet with state-of-the-art cross-media retrieval methods, including MHTN \cite{huang2018mhtn}, ACMR \cite{wang2017adversarial}, JRL \cite{zhai2014learning}, GSPH \cite{mandal2017generalized}, CMDN \cite{peng2016cross}, SCAN \cite{lee2018stacked}, GXN \cite{gu2018look}.
MHTN \cite{huang2018mhtn} learns common representations for 5 media types by transferring knowledge from single-media source domain (image) to cross-media target domain.
ACMR \cite{wang2017adversarial} learns the common representations by adversarial learning.
JRL \cite{zhai2014learning} applies semi-supervised regularization and sparse regularization to learn the common representations.
GSPH \cite{mandal2017generalized} proposes a generalized hashing method to preserve the semantic distance between two media types.
CMDN \cite{peng2016cross} first learns separate representations for each media by multiple deep networks, and then generates the common representations by a stacked network.
SCAN \cite{lee2018stacked} considers the latent alignments between image regions and text words to learn the image-text similarity. 
GXN \cite{gu2018look} incorporates generative processes into feature embedding to learn common representations.
Since SCAN and GXN are specially designed for cross-media retrieval between image and text, which are not easy to extend to cross-media retrieval among 4 media types. So we only compare with them on cross-media retrieval between image and text, and compare with the other state-of-the-art methods on cross-media retrieval among 4 media types. 

\subsection{Comparisons with State-of-the-art Methods}
In the experiments, we conduct two types of retrieval tasks, namely bi-modality fine-grained cross-media retrieval and multi-modal fine-grained cross-media retrieval. 
The results of our FGCrossNet and compared methods are shown in Tables \ref{expall} to \ref{exptoall}.

For the compared methods, we take the same features as the inputs for fair comparison. For image and video, if the input is not the original image, we take 200-dimensional CNN feature as the input, which is extracted from the FC layer of ResNet50. It is noted that the ResNet50 is fine-tuned on the image data of the new benchmark. For text, we take the 1,000-dimensional BoW feature as the input. For audio, we take the 128-dimensional MFCC features as the input.

\subsubsection{\textbf{Comparisons on Bi-modality Fine-grained Cross-media retrieval}}
\indent

Table \ref{expall} shows the MAP scores of bi-modality fine-grained cross-media retrieval of our FGCrossNet compared with existing methods on all 4 media types. We can see that our FGCrossNet achieves the best retrieval performance than all compared methods. Among the compared methods, MHTN achieves the best performance, which mainly because of its transfer learning ability from external single-media data to the cross-media data. But, our FGCrossNet achieves higher MAP scores than MHTN on all 12 bi-modality fine-grained cross-media retrieval tasks. It is mainly because of:
(1) Our FGCrossNet also applies transfer mechanism among cross-media data, which transfers the knowledge from image to text, video and audio. It is different from MHTN, which transfers the external knowledge to cross-media data.
(2) We design a uniform deep model to learn the 4 media data simultaneously, which mainly has the similar input forms and outputs, and reduces the heterogeneity gap to some extent.
(3) We jointly consider the classification constraint, center constraint and ranking constraint, to minimize the intra-class variance and maximize the inter-class variance.

SCAN adopts Faster R-CNN \cite{ren2015faster} to exploit the regions corresponding to objects, which is not suited for our new benchmark, due to that the images mostly only have one object.
GXN utilizes the generative models to improve the common representation learning. It is not suited for our new benchmark, since the image does not have a corresponding text descriptions. The text data in the new benchmark mainly focus on introducing the subcategory, not describing each image. So their performances on the new fine-grained cross-media retrieval benchmark are not well.

\subsubsection{\textbf{Comparisons on Multi-modality Fine-grained Cross-media retrieval}}
\indent

We also conduct the multi-modality fine-grained cross-media retrieval to verify the effectiveness of our FGCrossNet, and the results are shown in Table \ref{exptoall}. The trend is same with bi-modality fine-grained cross-media retrieval, our FGCrossNet achieves the best retrieval performance.
It is noted that our FGCrossNet has an advantage on the common representation learning of 4 media types, that it is a uniform and simple deep model, and can generate features for image, text, video and audio simultaneously. 
Among the compared methods, only MHTN can learn the common representations of 4 media types at the same time.
Although MHTN can learn the common representations of 4 media types simultaneously, its model is complex, where each media type has a special designed network stream.
For the other compared methods, they learn the common representations between 2 modalities, which increases the complexities of both training and testing. Taking I$\to$All as an example, we first conduct the bi-modality fine-grained cross-media retrievals between 2 modalities with their corresponding common representation, i.e. I$\to$T, I$\to$V and I$\to$A. Then we combine their results and the results of I$\to$I as the final results of I$\to$All.

\subsection{Ablation Study}
To verify the effect of each constraint in our FGCrossNet, we conduct ablation studies. The results are shown in Table \ref{ablation}. We can observe that: (1)
``Classification Constraint'' denotes that only using classification constraint to train our FGCrossNet, which also achieves better retrieval performance than all the compared methods. This shows that the classification constraint can help FGCrossNet learn the discriminative features that can discriminate similar subcategories.
(2)
``+Center Constraint'' denotes additionally using center constraint at the basis of classification constraint. It achieves the better retrieval performance than only using classification constraint by 0.043, which is because that center constraint enforces the features of the same subcategory clustering into its subcategory center.
(3)
``+Ranking Constraint'' denotes that applying all the three constraints in the loss function of FGCrossNet, which achieves the best performance except on ``I $\to$ A'' and ``A $\to$ I''. Ranking constraints focuses on the distinction between features of different subcategories, which boosts the retrieval performance.

\section{Conclusion}
\label{conclusion}
In this paper, we have made two highlights:
(1) \emph{A new benchmark} - We have constructed a new benchmark for fine-grained cross-media retrieval, which is the largest benchmark with the most media types and richest fine-grained subcategories. It will encourage further researches on fine-grained cross-media retrieval.
(2) \emph{A new approach} - We have proposed the FGCrossNet, which is a uniform deep model to simultaneously learn the common representations of 4 types of media.
We have jointly considered classification constraint, center constraint and ranking constraint to guide the common representation learning.
Extensive experiments verify the usefulness of the proposed new benchmark and effectiveness of our FGCrossNet.

The future work will lie in two aspects: (1) \emph{Task extension} - We will further explore the probabilities of other tasks, such as categorization and reasoning.
(2) \emph{Knowledge transfer} - In this paper, we have found that the performance of image or video is obviously better, we will focus on how to transfer the knowledge of image or video to text and audio for better retrieval performance.

\section*{Acknowledgment}
This work was supported by the National Natural Science Foundation of China under Grant 61771025.

%
\balance{
\bibliographystyle{ACM-Reference-Format}
\bibliography{sample-base}
}

\end{document}